\documentclass{aa}
\usepackage{graphicx,color}
\usepackage{natbib}
\usepackage{url}
\usepackage{amsmath}
\usepackage{amssymb}
\usepackage{longtable}

\bibpunct{(}{)}{;}{a}{}{,} % to follow the A&A style
\newcommand{\kms}{km\,s$^{-1}$}
\newcommand{\reff}{r$_\mathrm{eff}$}
\newcommand{\aeff}{a$_\mathrm{eff}$}

\begin{document}
 \title{Kinematics and stellar population of NGC 4486A}

  \author{Philippe Prugniel,
          \inst{1}
 \and
 Werner Zeilinger
 \inst{2}
 \and
 Mina Koleva
 \inst{3,1}
 \and
 Sven de Rijcke
 \inst{4}
        }

   \offprints{Ph. Prugniel}

   \institute{Universit\'e de Lyon, Universit\'e Lyon 1, Villeurbanne, F-69622, France;
              CRAL, Observatoire de Lyon, CNRS UMR~5574, 69561 Saint-Genis Laval, France\\
              \email{philippe.prugniel@univ-lyon1.fr}
              \and
  Institut  f\"{u}r Astronomie, Universit\"{a}t Wien, T\"{u}rkenschanzstra{\ss}e
  17, A-1180 Wien, Austria 
              \email{werner.zeilinger@univie.ac.at}
              \and
  Instituto de Astrof\'{\i}sica de Canarias, La Laguna, E-38200 Tenerife, Spain; Departamento de Astrof\'{\i}sica, Universidad de La Laguna, E-38205 La
Laguna, Tenerife, Spain\\
              \email{koleva@iac.es}
              \and
  Dept. of Physics \& Astronomy, Ghent University, Krijgslaan 281, S9, B-9000 Ghent, Belgium
              \email{sven.derijcke@ugent.be}
 }

   \date{Received ; accepted }

% \abstract{}{}{}{}{}
% 5 {} token are mandatory

  \abstract
   %context heading (optional)
{
NGC\,4486A is a low-luminosity elliptical galaxy harbouring an
edge-on nuclear disk of stars and dust. It is known to host a super-massive
black hole.}
  {
We study its large-scale kinematics and stellar population along the major 
axis to investigate the link between the nuclear and global properties.
}
    {
We use long-slit medium-resolution optical spectra that we fit against
stellar population models.
     }
     {
The SSP-equivalent age is about 12 Gyr old throughout the body
of the galaxy, and its metallicity decreases from [Fe/H] = 0.18
near the centre to sub-solar values in the outskirts. 
The metallicity
gradient is $-0.24$ dex per decade of radius within the effective isophote.
The velocity dispersion is $132\pm3$~\kms{} at 1.3 \arcsec{} from the centre
and decreases outwards. The rotation velocity reaches a maximum 
 $V_{\rm max} \gtrapprox 115\pm5$ \kms{} at a radius  $1.3 < r_{\rm max} < 2$ \arcsec.
}
   {
NGC\,4486A appears to be a typical low-luminosity elliptical galaxy.
There is no signature in the stellar population 
of the possible ancient accretion/merging event 
that produced the disk. 
}
   \keywords{galaxies: individual: NGC4486A - galaxies: elliptical and lenticular, cD - 
   galaxies: kinematics and dynamics -  galaxies: stellar content}
 \authorrunning{Prugniel et al.}
   \titlerunning{NGC\,4486A}

  \maketitle
%
%________________________________________________________________

\section{Introduction}

\object{NGC\,4486A} is a low-luminosity (M$_B = -17.77$ mag) E2 galaxy belonging
to the Virgo cluster.
Its catalogue properties are summarized in Table~\ref{tab:properties}.
It is one of the four galaxies projected within a few
arcmin of the central cD, \object{M\,87} \citep{prugniel1987}.
It habours a spectacular nuclear disk of stars and dust
\citep{kormendy2005}.
This almost edge-on disk, detected up to a radius of 250-350 pc 
(3--4 \arcsec, approximately half the effective radius, \reff),
is marginally bluer than the surrounding population, suggesting that it
is at least 2 Gyr younger.
It is aligned along the major axis of the galaxy.
This disk is believed to have been formed from accreted gas that
funneled down to the central region.

Although it is a bright galaxy, it is not often observed because of a
foreground bright star located 2.5 \arcsec{} away from the nucleus
\citep{rc1,devauc1959}.  However, this particular circumstance makes
this object a first choice target for adaptive optics observations.
\citet{kormendy2005} used the adaptive optics system of the CFH telescope
to reach a spatial resolution of 0.07 \arcsec{} (FWHM) after a Lucy
deconvolution.  \citet{nowak2007} used the near-infrared integral
field spectrograph SINFONI to study the central kinematics.  Using the
Schwarzschild orbit superposition method to fit the two-dimensional
information, they found a central super-massive black hole of mass $
M_\bullet = 1.25 \times 10^7$ M$_\odot$ and rejected any model without
a black hole at the 4.5 $\sigma$ confidence level.

In this paper, we study the internal kinematics and the stellar
population of this galaxy to investigate if its nuclear
particularity is related to its global properties.

The paper is organized as follows: in Sect. 2 we present the
observations and the data reduction, in Sect. 3 the analysis, and
in Sect. 4 we discuss the results and conclude.

\begin{table*}
\centering
\caption{\label{tab:properties}Characteristics of NGC\,4486A.}
\begin{tabular}{l|c|c}
\hline\hline

Characteristics                            & Value               & Ref.\\
\hline
Classification                             & E2                    & 1 \\
Asymptotic apparent magnitude              & V$_T$ = 12.53         & 1 \\
Distance modulus                           & 31.31 mag             & 2\\
Systemic radial velocity                   & $757\pm6$ \kms        & 0\\
Absolute V band magnitude\tablefootmark{a} & -18.86 mag & 1\\
Effective radius                           & 7.4 \arcsec{} = 650 pc   & 1\\
Mean surface brightness                    & $\mu_{\rm eV}$ = 19.63 mag & 1\\
Ellipticity                                & $\epsilon(r_e)$ = 0.25 & 1\\
Position angle (N to E)                    & 22 deg & 0\\
S\'ersic index                             & 2.04                  & 1\\
Central velocity dispersion                & 110~\kms& 3\\
Maximum rotation velocity                  & 115~\kms& 0\\
Age                                & 12 Gyr & 0\\
Central metallicity                        & $0.18\pm0.03$ dex & 0\\
Metallicity gradient                       & $\nabla_{\rm [Fe/H]} = 0.24$& 0\\

\hline
\end{tabular}
\tablefoot{ 
The references are: (0) This paper; (1) \citet{kormendy2009};
(2) \citet{mei2007}; (3) \citet{nowak2007}\\
\tablefoottext{a}{
Corrected for Galactic extinction using \citet{schlegel1998}}
}
\end{table*}

\begin{table}
  \caption{\label{tab:setup}
Setup and journal of the observations.}
  \begin{tabular}{lcc}
    \hline
\hline
\multicolumn{2}{c}{Gemini-South program GS-2008A-Q-3}\\
\hline
CCDs, EEV\#             & 2037-06-03/8194-19-04/8261-07-04 \\
\# of pixels            & 3 $\times$ 2048$\times$4068 chips        \\
pixel size [$\mu$m$^2$] & 13.5$\times$13.5  \\
scale [arcsec pix$^{-1}$]  & 0.0727 \\
readout noise [e$^-$ pix$^{-1}$] & 3.20 / 3.50 / 3.10    \\
gain [ADU (e$^-$)$^{-1}$]        & 2.000 / 1.900 / 1.900 \\
grism                      & B600\_G5323  \\
slit width [\arcsec]        & 0.5         \\
slit position angle [deg]  & 18         \\
FWHM $\delta\lambda$ [\AA]     & 2.5     \\  
$\sigma_{\rm instr}$ [km s$^{-1}$] & 61      \\
dispersion [\AA\,pix$^{-1}$]   & 0.46    \\
\hline
\multicolumn{2}{c}{First setup: 2008 Apr. 2}\\
\hline
central wavelength  [\AA]          & 5250 \\
spectral range [\AA]       & 3640 -- 6500 \\
exposures [s]                & 4 $\times$ 1200\\
seeing [\arcsec]              & 0.9 \\
\hline
\multicolumn{2}{c}{Second setup: 2008 March 13/14}\\
\hline
central wavelength [\AA]              & 5300 \\
spectral range [\AA]       & 3640 -- 6500 \\
exposures [s]                &1500 + 2 $\times$ 1888\\
seeing [\arcsec]              & 1.6 \\
\hline
  \end{tabular}
\end{table}

\section{Observations and data reduction}

In the frame of another project \citep{koleva2011}, we discovered
observations of NGC\,4486A in the Gemini Science Archive.
They were taken with the GMOS spectrograph attached to the 8.1 m 
Gemini-South telescope in long-slit mode.
The setup and journal of these observations are reported in 
Table~\ref{tab:setup}.
Two slightly different grating orientations were used 
to patch the holes in the wavelength coverage, which are caused by the
separation between the three CCD detectors.

\begin{figure}
\centering
\includegraphics[width=0.4\textwidth]{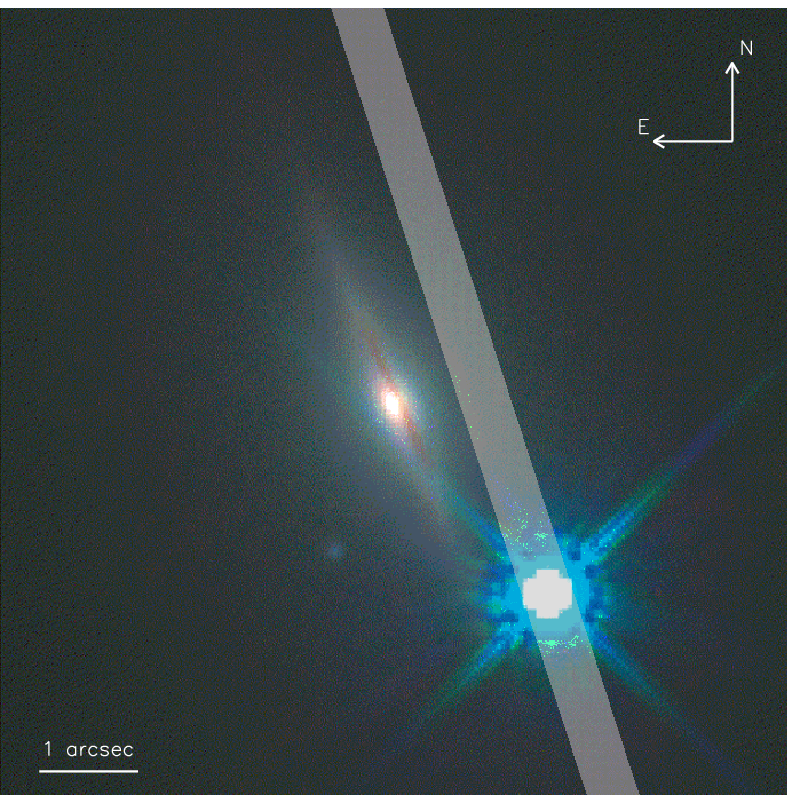}
\caption{Central field of NGC\,4486A.  
  The location of the slit is superimposed on the composite
image from \citet[their Fig. 4]{kormendy2005}.
The size is 8 \arcsec{} across.
}
\hskip 0cm
\label{fig:slit}
\end{figure}

The observer probably ignored the presence of the star and inadvertently
centred the slit on it, unfortunately missing the centre of
the galaxy by about 1 \arcsec{} (Fig.~\ref{fig:slit}).
Nevertheless, these spectra provide a chance to probe the stellar 
population and the kinematics of this object.

The data reduction was made using the GMOS pipeline in the IRAF
environment exactly as described in \citet{koleva2011}.  The
instrumental broadening, or line-spread function (LSF), was found to be
acceptably modelled with a Gaussian with standard deviation
$\sigma_{\rm ins} = 61$~\kms.  Some pixels within $r = 0.6$ \arcsec{} from
the peak of light are saturated. The seeing measured on images taken
during the same observing blocks is for the two nights 0.9 and 1.6~\arcsec{} 
FWHM with a typical error of $\pm 0.06$ \arcsec.

The radial position, $r$, of a spectrum with respect to the galaxy's 
kinematical centre
is related to the position along the slit with respect to the peak of
light, $x$, as: $r^2 = (x+2.15\pm0.07)^2 +  (1.28\pm0.07)^2$ \arcsec$^2$.

In order to measure the distance of the galaxy nucleus orthogonal to
the slit, an ACS image of NGC\,4486A obtained in the F850LP filter was
extracted from the Hubble Legacy Archive. The distance between the
star and the nucleus was determined to be $2.51\pm0.05$ \arcsec{} from
measuring the position of the respective peaks of intensity.
According to this measurement, the light peak of the galaxy in the
slit should be at $2.36$ \arcsec{} from the star.  Fitting two
Gaussians to the light distribution of the galaxy nucleus and the star
in the GMOS acquisition image, which has been taken before the
spectra, we measure a corresponding separation of $2.35\pm0.08$ \arcsec.
Using the light distribution of the spectrum along the slit, and
fitting two Gaussians to the light peaks, we measure a separation of
$2.45\pm0.1$.

\citet{nowak2007} did not report any displacement of the kinematical centre,
and therefore, we will not consider the small discrepancy between our photometric and
kinematical determinations (one pixel, or one fifth of the FWHM seeing)
as significant; it may result from
the dust extinction.
We adopt the kinematical centre as a reference.

\section{Data analysis}

The analysis was performed with the full-spectrum fitting package 
ULySS\footnote{http://ulyss.univ-lyon1.fr} \citep{ulyss}.

The present difficulty of contamination by a foreground or background
object was already met on other occasions. In \cite{bouchard2010} and
\citet{makarova2010}, the interloper was separately modelled and then
subtracted before studying the target. In the present case, the
contamination reaches 99\% and extends over a large range of the
object. We therefore tried to optimise the decomposition.

The ULySS package allows one to fit a spectrum against a constrained
linear combination of non-linear models. In the present case we are
considering two components that represent the star and the
galaxy respectively.  For the `star' component we could have attempted to use a
spectrum extracted in the region where the galaxy spectrum is
quasi-negligible.  But we found a simpler and more robust way to
model the stellar spectrum with ULySS, as in \citet{wu2011}.  In
this paper, stellar spectra are fitted against empirical models
consisting of interpolated ELODIE spectra, and the authors
show that this method reproduces accurately any observation and
restores reliable atmospheric parameters.

The adopted model is therefore

\begin{align}
{\rm Obs}(\lambda, x) = &P_n(\lambda)  \times {\rm LSF}~\otimes  \nonumber\\
&\bigg(
(1-f) \times S(T_{\rm eff}, g, \mathrm{[Fe/H]}_s, \lambda)~+\nonumber\\
&f \times G(cz, \sigma, h_3, h_4, {\rm Age}, \mathrm{[Fe/H]}_g, x, \lambda)\bigg),
\label{eqn:main}
\end{align}

where ${\rm Obs}(\lambda, x)$ is the observed long-slit spectrum,
function of the wavelength, $\lambda$, and position along the slit,
$x$.  $S$ is the spectrum representing the foreground star, function
of the atmospheric parameters temperature, surface gravity and
metallicity, and $G$ is a synthetic stellar population, function of
the radial velocity, velocity dispersion, $h_3$ and $h_4$ coefficients
of the Gauss-Hermite expansion, age and metallicity for the position
$x$ along the slit.  $f$ is the relative light fraction of $G$, LSF
is the line-spread function, and $P_n$ is a Legendre multiplicative
polynomial of degree $n$.

The stellar model, $S$, is an interpolated empirical spectrum
computed using the ELODIE library \citep{ELODIE31,PS01}.
The population model, $G$, is computed with PEGASE.HR \citep{PEGASEHR}
and is also based on the ELODIE library.
We are using single stellar populations (SSPs) with a \citet{salpeter1955}
IMF.
An SSP supposes that all the stars are coeval and have the
same metallicity.
Because the two components $S$ and $G$ are based on the same spectral
library, they share the same LSF (resolution R~$\approx$~10000),
and therefore the LSF can be factorized.
The multiplicative polynomial, $P_n$, is modelling the shape of the
spectrum and makes the flux calibration unnecessary.
It was shown that even with flux-calibrated spectra, this polynomial
is an advantage to match the inaccuracies of this calibration. We used a
degree $n = 40$, determined as described in \citet{ulyss} which was also
adopted in \citet{koleva2011}.

Because the stellar model does not depend on the position along the
slit, it appears preferable to determine its parameters and to freeze
them when analysing the galaxy. Therefore, the free parameters for the
galactic fit are $cz$, $\sigma$, Age, ${\rm [Fe/H]}_g$, $f$ and the
coefficients of $P_n$.  There is only one more parameter than for a
 SSP fit.  We discuss below the adjustment of the stellar
model and then the fit of the galaxy.

\subsection{Fit of the stellar spectrum}
\label{sect:star}

To determine the best ELODIE model to represent the foreground star,
we first ran the analysis with the model of Eq.~\ref{eqn:main},
leaving the stellar and galactic parameters free (as well as $P_n$).
The retrieved parameters were very stable (i. e. independent of the
position in the central two arcsec).  Excluding the central spectra,
because some pixels are saturated, we adopted the following solution:
$T_{\rm eff} = 6335\pm5$\,K, log (g, g/cm$^2$) $= 4.14\pm0.03$ and
[Fe/H] $ = -0.33\pm0.02$.  The quoted uncertainties reflect the
dispersion of the values for the different extracted spectra (the
physical uncertainties are larger, see \citealt{wu2011}).

\begin{figure}
\centering
\includegraphics[width=0.4\textwidth]{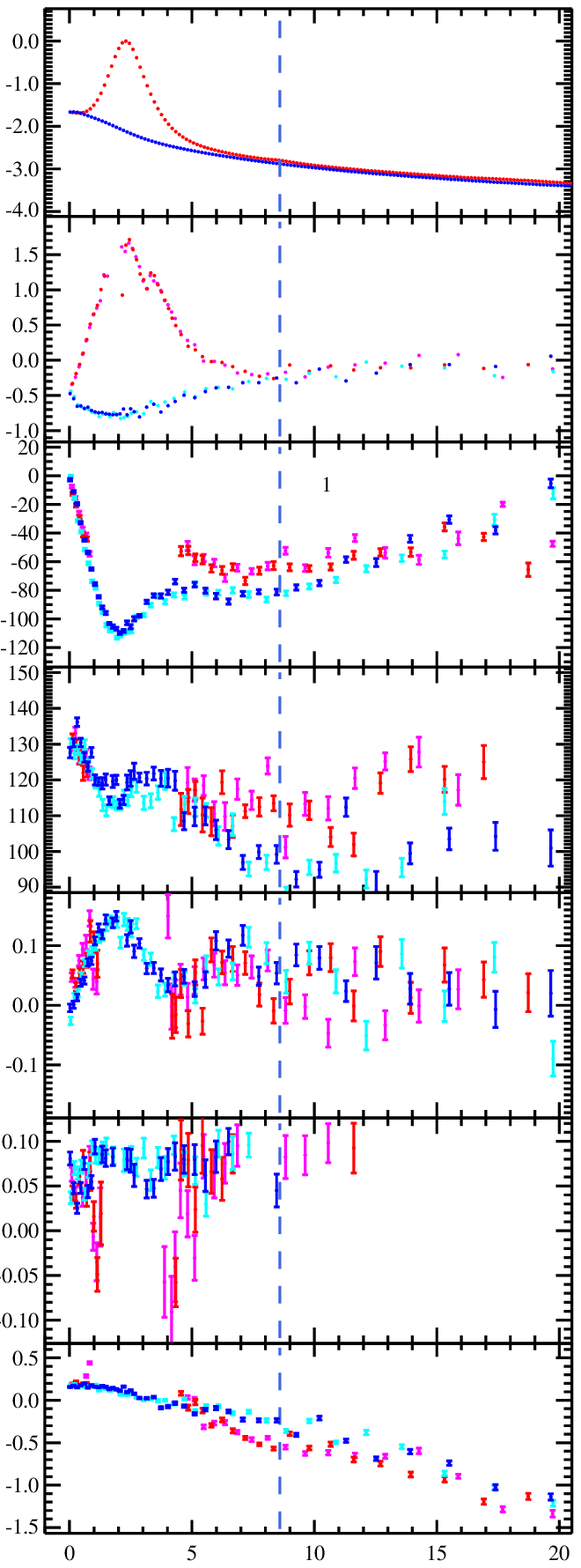}
\caption{Radial profiles of NGC\,4486A derived from SSP fits.  The
  spectra are radially binned for a compromise between S/N and spatial
  resolution (they have at minimum S/N = 40).  The profiles are folded
  around the position of the kinematical centre.  The red points are
  for the positive radii (South side) and blue for negative; the tones
  separate the two setups (the darker colours are for the first setup).  
  The abscissa is the position along
  the slit ($a = x - 2.15$). The vertical blue dashed line marks the effective
  isophote.
  From top to bottom the panels are: (1)
  The logarithm of the relative total flux. (2) The logarithm of
  1/f. (3 to 7) The internal kinematics and the SSP-equivalent
  metallicity.
  The velocity and $h_3$ of the north side (blue points) are
  symmetrized. The points with $f< 0.5$ were excluded.  }
\hskip 0cm
\label{fig:profile}
\end{figure}

\begin{figure*}
\centering
\includegraphics[]{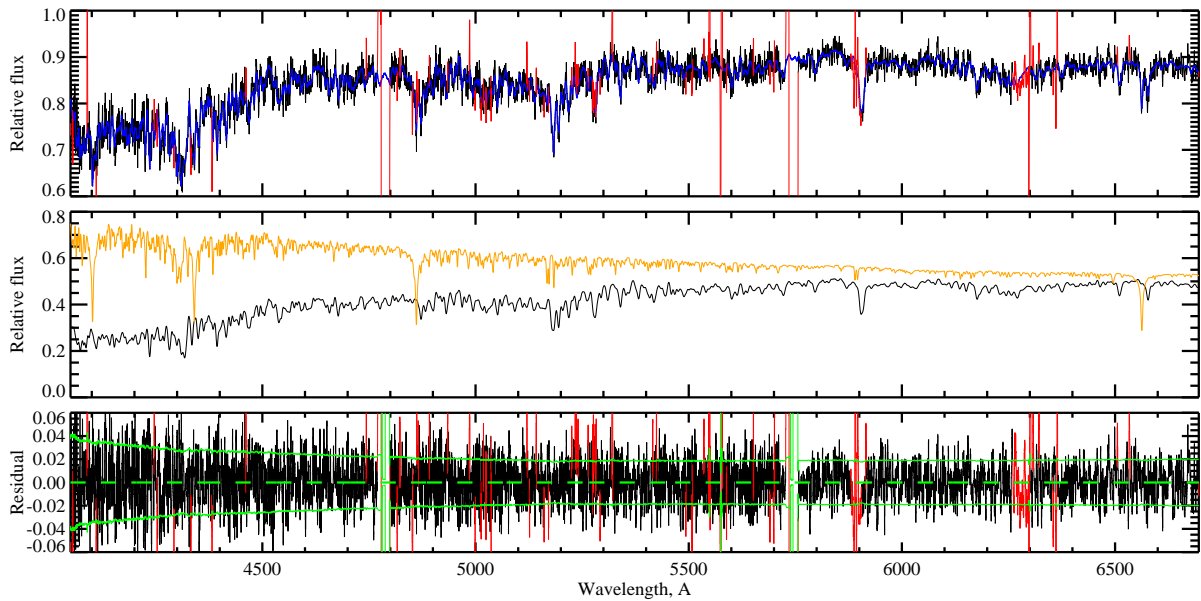}
\caption{Fit of a spectrum of NGC\,4486A.  
The top panel shows the observed (in black) and fitted (blue) spectra.
The red points mark the excluded regions (because of strong telluric
absorption) and the rejected pixels (kappa-sigma clipping). The
two gaps between the detectors are visible near 4800 and 5750~\AA.
The middle panel shows the two components of the model separately,
the galactic population in black, and the star in orange.
The bottom panel presents the residuals (observation - model).
The green lines are the 1-$\sigma$ errors.
}
\hskip 0cm
\label{fig:fit}
\end{figure*}

\subsection{Kinematics and population of NGC\,4486A}

In order to check if this decomposition can bias the results, we applied
the same analysis on another galaxy of the same observing run, which was
un-affected by a star. We used \object{NGC\,4434} because it has approximately
the same diameter, velocity dispersion, age, and metallicity as our present
target. 
We found that the contribution of the stellar spectrum amounts to
approximately 2\%, which means that the flux contribution is comparable to
the rms noise (the spectra were rebinned  to reach
S/N=40). 
Because this contribution is constrained to be positive,
we expected this component to be often inactive.
Surprisingly, it is the case for only 3\% of the individual 
spectra, independently of the radius of these extractions.
These fake detections of the
stellar component have no effect on the determination of the kinematics
or of the age, but a small bias of 0.03 dex on the metallicity.
This bias is acceptable for the present purpose.

We repeated the analysis of NGC\,4486A using the atmospheric parameters 
determined in Sect.~\ref{sect:star}.  
The profiles obtained with the two setups are perfectly
coincident, and the kinematics on the two sides of the galaxies fold
adequately.  The age is constant at 12 Gyr.  The central metallicity
is nearly solar, ${\rm [Fe/H]} = 0.05\pm0.02$~dex, but the profile is
asymmetric, declining to -0.65 on the north side and to -1 dex on the
south at a radius of 10~\arcsec.  The light of the star exceeds the one
from the galaxy by two orders of magnitude at the location of the star
(the galactic spectrum is negligible at this place). The contamination
then decreases, but remains at a level higher than $15$\% over the
whole radial range.

This asymmetry of the metallicity is probably not physical, because the
galaxy appears regular in other respects and this asymmetry is not
observed in other galaxies. Therefore, we carefully examinated the
fits and did additional tests.  We separately analysed the blue and
the red halves of the spectra and found consistent results (though
naturally more noisy; the
blue half was below 5000 \AA, and the red above 4800~\AA).
When selecting the red region from 4900~\AA{} (excluding H$\beta$),
we obtained a marginal reduction of the asymmetry.
We also tested with no more success, if allowing for
a differential extinction between the star and the galaxy would reduce
the effect and improve the fit.

The close examination of the fits revealed the probable presence
of additional continuum light, in particular on the south side of the
galaxy, which is more affected by the star. This pushes the solution to a lower
metallicity and degrades the quality of the fit.

We then compared the light profiles along the slit for our target and
for a template star observed with the same setup, and found that
the contamination is actually higher than the one inferred from the
spectroscopic decomposition. This provides additional evidence for
pollution by diffuse light, which was not (or not perfectly) dispersed.
The scattered light represents 1.5\% of the incident light.
Unfortunately, we did not have adequate observations to check if the
PSF before the slit has indeed lower wings (the acquisition images
are saturated).

Based on this suspicion of diffuse light, we fitted models with an
additional free additive term. However, such a component is naturally
degenerate with the metallicity, and the fit became unstable. It
returned on average higher and super-solar ${\rm [Fe/H]}_g$ (as expected).
We therefore added constraints.

We modified the model of Equation~\ref{eqn:main} by replacing the
stellar component, $S$, by a combination $(1-a) \times S + a \times
C$, where $C$ is a constant spectrum representing the undispersed
light and $a$ the corresponding fraction.  We adjusted $a$ as the
function of the distance to the star along the slit to match
the retrieved spectroscopic contamination, $f$, with the photometric
contamination obtained comparing our observation with one of a
template star.  We tried if using a different spectral energy
distribution in place of the constant $C$ affected the result and the
quality of the fit. We considered either the same distribution as the
star (using a smooth spectrum of it) or a spectrum biased to the blue
(because the undispersed light may contaminate more the blue part of the
spectrum).  But we found no significant difference between these choices and
adopted the additive constant.  Because the age was apparently
homogeneous, we fixed it to 12 Gyr to reduce the degree of
freedom.

The final profiles of the kinematics and of the stellar population are
presented in Fig.~\ref{fig:profile}.  We rejected the points in the
range  $-1.3 < x < 1.9$~\arcsec  ($1.5 < r < 4.2$~\arcsec on the
southern side), where the contamination from the foreground star
exceeds the galactic light by more than a factor 2 .
An example of the decomposition is shown on 
Fig.~\ref{fig:fit} for the spectrum at the location $x=3, 
r=5.6$~\arcsec, where the two contributions are comparable.
The residuals are consistent with the noise computed from the photon 
statistics.

The kinematical, and now the metallicity, profiles are symmetric.  The
systemic velocity is $757\pm6$ \kms.  The peak of rotation velocity, $
V_{\rm max} = 107\pm3$, is reached at 2~\arcsec{} along the major axis
(i. e. at a galactocentric distance of $r = 2.4$~\arcsec).  The
velocity dispersion in our most central location,  at 1.3 \arcsec{} along
the minor axis, is $\sigma_0 = 132\pm3$~\kms.
Assuming a Gaussian line-of-sight velocity distribution (LOSVD),
the peak rotation velocity is  96 \kms{} and the velocity dispersion
 $\sigma_0 = 139\pm3$~\kms. The difference is due to the h3 and h4
moments.
The h3 profile reveals a strong asymmetry of the LOSVD
strongly correlated with the peak of rotation.
After the peak, the rotation velocity and the velocity dispersion
linearly decrease until 20 \arcsec. The rotation decreases to 35~\kms{}.
The metallicity decreases
linearly with radius from the central [Fe/H] = 0.18 dex to -1.3 dex 
at 20 \arcsec.

Our analysis actually extends further than the 20 \arcsec{} limit 
of Fig.~\ref{fig:profile} (2.3~\aeff),
but owing to the lower S/N and strong contamination we have less
confidence in these results. The radial velocity and metallicity appear
well fitted, as the regularity and symmetry of the profiles show, but
$\sigma$ and the higher moments cannot be reliably determined. There
is a hint that the rotation velocity continues to decrease and reaches 0 at
25-30 \arcsec, but no indication that the outskirts would rotate in the
opposite direction.  This halo may be non-rotating. The metallicity continues
to decrease to $-2$ dex at 30 \arcsec, and stays constant afterwards.
However, the metallicity is the parameter that is the most affected by the 
contamination, and we will not interpret these external points.

\begin{figure}
\centering
\includegraphics[width=0.45\textwidth]{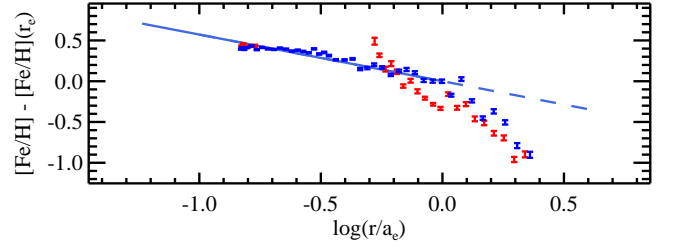}
\caption{Fit of the metallicity gradient. The colours are as in Fig.~\ref{fig:profile}. The abscissa is the distance to the galaxy centre, divided by the effective semi-major axis. The straight line is the solution of the power-law fit, and its continuous region marks the used radial range.
}
\hskip 0cm
\label{fig:grad}
\end{figure}

We fitted the metallicity gradients, $\nabla_{\rm [Fe/H]} = \Delta
{\rm [Fe/H]} / \Delta \log(r)$, using the points between the innermost
location (r = 1.3 \arcsec, greater than the seeing disk) and the
effective isophote.  The fit is shown in Fig.~\ref{fig:grad} and the
value in Table~\ref{tab:properties}.  Clearly the adopted
power-law model is not well suited (as in some other
cases, see \citealt{koleva2011}), but this fit allows us to compare this
object with other galaxies. The flattening of the profiles in the most
central region may be caused by the fact that the observation is made
parallel to the major axis. The steepening after one \reff{},
though consistent between the two sides and between the two setups,
should be regarded with some caution, because the metallicity is the most
sensitive parameter with respect to the pollution by diffuse light.

\section{Discussion and conclusion}

\begin{figure}
\centering
\includegraphics[width=0.45\textwidth]{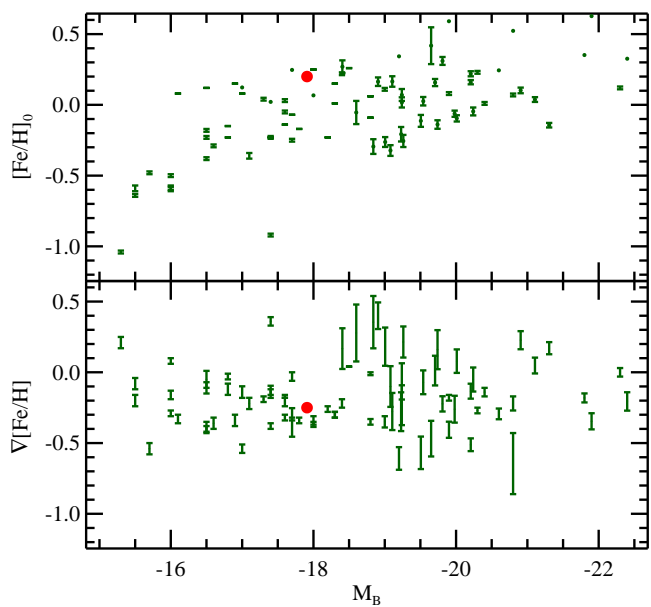}
\caption{Location of NGC\,4486A in the metallicity and the metallicity gradients vs. luminosity diagrams. NGC\,4486A is shown as red dots, and the other points are the galaxies from \citet{koleva2011}.
}
\hskip 0cm
\label{fig:scal}
\end{figure}

\subsection{Revision of the catalogue properties}

Because of the foreground star, the radial velocity and the velocity
dispersion reported in the databases (HyperLeda and NED) are wrong.
The velocity $\mathrm{cz} = 150$~\kms{} and the dispersion $\sigma_0 =
41\pm5$~\kms{} obtained by the ENEAR project \citep{wegner2003}
correspond to the stellar spectrum. The other earlier measurements are
affected by the same problem.

\subsection{Comparison with Nowak et al.}

\citet{nowak2007} find a kinematically cold component in the centre, with a
velocity dispersion dropping to about $\sigma_0 = 110$ \kms. This structure 
coincides with the disk, and one arcsec above it, the velocity dispersion
is in the range 130 to 140 \kms.
This is fully consistent with our value at the same location.
The same method, using the near-infrared CO band heads, was used by 
\citet{nowak2008} to study
\object{NGC\,1316}, which was also studied by us in the optical domain \citep{koleva2011}.
Our value of the velocity dispersion was significantly higher
than the one of \citet{nowak2008} (238$\pm2$ vs. 221 to 226 \kms). 
Because our measurements were based on
high S/N observations at a high spectral resolution ($\sigma_{\rm ins} = 19$ \kms),
we tend to trust them. 
The usage of the CO band heads to measure the velocity dispersion
may be sensitive to metallicity effects, or, because of the significant
dust extinction, the stars probed by the NIR observations may be different
from those probed in the visible.
Anyway, in the present case, the two methods agree to perfection.

\citet{nowak2007} find that the velocity increases up to the edge of their field
at 1.3 \arcsec. At this point they note a rotation velocity of
$115\pm5$ \kms. We observe the peak at 2 \arcsec along the major axis, 
which is certainly 
consistent, given the difference of spatial resolution.
Therefore, the peak of the rotation is reached between 1.3 and 2 \arcsec,
and the maximum rotation velocity is $V_{\rm max} \gtrapprox 115\pm5$ \kms. The 
rotation velocity decreases monotically outwards.

\subsection{NGC\,4486A, a typical low-luminiosity elliptical}

Figure~\ref{fig:scal} locates NGC\,4486A in the metallicity  and metallicity gradients vs. 
luminosity diagrams.
The other galaxies represented in this figure are from \citet{koleva2011}.

NGC\,4486A has a uniformly old population, a super-solar
central metallicity, and a negative metallicity gradient
$\nabla_{\rm [Fe/H]} = -0.24$. 
These properties are similar to other galaxies in the
same mass range.
The orientation of the nuclear disk suggests that the galaxy is observed 
edge-on. The considerable rotation, implying $V_{\rm max}/\sigma_0 \approx 1$, is
produced by a central concentration of mass and does not reflect a
significant angular momentum.

The high h3 value, coinciding with the peak of rotation, indicates
a population of fast rotating stars, and the steep decrease of
the rotation in the external regions is consistent with the hypothesis
proposed by \citet{kormendy2005} for the formation of the disk.
An ancient merger, 5 to 10 Gyr ago (because we do not see the effect on the population), 
injected some gas that cooled to the centre and formed a disk where it formed 
stars. This corresponds to the population observed in rotation, 
while the pre-existing population conserved a low rotation.
However, because there is no signature in the stellar population, this
scenario remains speculative.

\subsection{Summary}

We analysed long-slit spectra of NGC\,4486A taken along its major axis
and derived its internal kinematics, its age, and its metallicity.
It is a typical low-luminosity elliptical galaxy.

We showed that with a careful composite modelling and a full-spectrum
fitting it is possible to correct a galactic spectrum from the
contamination by an external source of light and derive consistent
information about the physical properties of the galaxy. 
The same approach can be used to separate a non-stellar emission in
the nuclear region.

\begin{acknowledgements}

PhP acknowledges the support from the Programe National Cosmologie et 
Galaxies (PNCG), INSU/CNRS.
PhP and WZ acknowlege a bilateral collaboration grant between France
and Austria (AMADEUS, collaboration project PHC19451XM).  SdR and PhP
acknowledge a bilateral collaboration grant between the Flander region
and France (Tournesol).
MK has been supported by the Programa 
Nacional de Astronom\'{\i}a y Astrof\'{\i}sica of the Spanish Ministry 
of Science and Innovation under grant \emph{AYA2007-67752-C03-01} and 
Bulgarian Scientific Research Fund \emph{DO 02-85/2008}.
She thanks CRAL, Observatoire de Lyon, Universit\'{e} Claude
Bernard for an Invited Professorship.
We thank Nina Nowak and the referee for comments on the manuscript.

\end{acknowledgements}

\bibliographystyle{aa} % style aa.bst
\bibliography{n4486a}  % bibtex database

\end{document}